\documentclass[a4paper,12pt]{article}
\usepackage{epsf,multicol,ifthen}
\usepackage[dvips]{graphicx}
\usepackage{cite}
\usepackage{hyperref}
\newcommand{\bec}{\begin{center}}
\newcommand{\ec}{\end{center}}
\newcommand{\bee}{\begin{equation}}
\newcommand{\ee}{\end{equation}}

\title{B-tagging and  searches for new physics beyond the Standard Model}

\author{T.V. Obikhod\thanks{E-mail: obikhod@kinr.kiev.ua}, I.A. Petrenko\\
\small\emph{Institute for Nuclear Research, National Academy of Science of Ukraine} \\ 
\small\emph{47, prosp. Nauki, Kiev, 03028, Ukraine}}

\begin{document}
\maketitle

\abstract{The article is devoted to the searches for new particles predicted by physics beyond the Standard Model through the b-tagging algorithm. The dependence of b-tagging efficiency on the  jet identification, impact parameter identification, secondary vertex identification, kinematic cuts is studied  with the help of computer programs Pythia  8.2 and  Fastjet 3.3.0.  The selection criteria for kinematic parameters, their ratios for an optimal result on the reconstruction of the vertices of heavy particles are found.} \\ 
\vspace*{3mm}

\newpage
\section{Introduction}

	The search for new physics beyond the Standard Model at the LHC is the most important task of modern high energy physics. The method for accomplishing this task is connected with the need of searches for heavy particles like superparticles, Kaluza-Klein particles and micro black holes. As LHC is a factory of top quarks, almost all new physical scenario that resolves the hierarchy problem will include new heavy particles with decay to top quarks. As top quark decays to a b quark and to the W boson decaying to two light quarks, therefore the top quark decay is connected with the appearance of three quarks in the calorimeter. Therefore, the ability of b-quark identification and recovering of new particles through the decay vertices is the part of the searches for new physics at the LHC. 

	The most important identification of particles towards an optimal determination of their direction, energy and type is the reconstruction of particle flow \cite{1.}. The list of individual particles is then presented as analog of Monte-Carlo event generator for building jets from the information of the quark and gluon energies and directions. This information is used for the reconstruction of the decay products which can be not only Standard Model (SM) particles but also new heavy particles predicted by supersymmetry or theories of extra dimensions (e.g. Randall-Sundrum or Arkani-Hamed-Dimopoulos-Dvali theories, etc.) and so on. 
	
		The fractions of  charged particles, photons and neutral hadrons (65\%, 25\% and 10\% respectively) ensure that 90\% of the jet energy can be reconstructed with the particle-flow algorithm. As a consequence, it is expected that jets made of reconstructed particles are much closer to jets made of Monte Carlo–generated particles than jets made from the sole calorimeter information. 
		
	The reconstruction of jets made of reconstructed particles is connected with the jet identification. The identification of bottom quarks called b-tagging is the most appropriate method of jet flavor tagging used in modern particle physics experiments. B-tagging sheds light on the following processes:

    $\bullet$ CP violation \cite{2.};

    $\bullet$ Formation of heavy particles, both recently discovered SM particles (B-mesons of special type) and expected in the future;

    $\bullet$ SUSY searches;
    
    $\bullet$ micro black hole searches;
    
    $\bullet$ Higgs boson decay into b quarks, \cite{3.}. A precise measurement of the Higgs boson decay, H$\rightarrow$bb, with formation of two b-jets, presented in \cite{4.}, directly probes the Yukawa coupling of the Higgs boson to a down-type quark. This decay is a test of the hypothesis that the Higgs boson is connected with mass generation in the fermion sector of the SM. 
    
    The paper is devoted to the searches for new physics beyond the SM with the help of b-tagging identification to the study its dependence on parameters. For disentanglement of the decay products of the exotic particles from the SM background, it is necessary to reconstruct as many final particles as possible. The particle identification is performed with a combination of the information from particle tracks, calorimeter clusters, and muon tracks. These constraints led to the advanced tracking and clustering algorithms. It is connected with the kinematic cuts on the decay products of pp-collisions, with the impact parameter and secondary vertex identification by using Fastjet 3.3.0 and Pythia 8.2 computer programs. The determination of optimal value of parameters for the effective recovery of the secondary vertex of heavy particles is the main aim of this paper. 
    
\section{B-tagging identification}
It is necessary to reconstruct as many final particles as possible for disentanglement of the decay products of the exotic particles from the SM background. 
B-tagging identification, connected with b-quark signatures,
has following features and benefits for experimental determination of primary particles:
 
    $\bullet$ hadrons containing b quarks have sufficient lifetime;
    
    $\bullet$ presence of secondary vertex (SV);
    
    $\bullet$ tracks with large impact parameter (IP);
    
    $\bullet$ the bottom quark is much more massive, with mass about 5 GeV, and thus its decay products have higher transverse momentum; 
    
    $\bullet$ b-jets have higher multiplicities and invariant masses;
    
    $\bullet$ The B-decay produces often leptons. 

	The cutting of the jet mass and the cut variables is associated with the $k_t$ algorithm with clustering each jet to search for subjets as follows:

	$\circ$ particles are clustered into jets of size R, when the pair is closest in 
$\Delta R = \sqrt{\Delta\eta^2 + \Delta\phi^2}$. This pair merges into a single four-vector, and then procedure repeats. The procedure ends when no two four-vectors have $\Delta R < R$;

	$\circ$ with these subjets are imposed additional kinematic cuts: jets must satisfy the absolute constraints, $p_T > 50$ GeV and $|\eta| < 2.5$ to be considered for analysis.

	The plot of a typical top jet with displayed energy, pseudorapidity, $\eta$, and azimuthal angle, $\phi$ is presented in fig. 1, from \cite{5.}.
\bec 
{\includegraphics[width=0.59\textwidth]{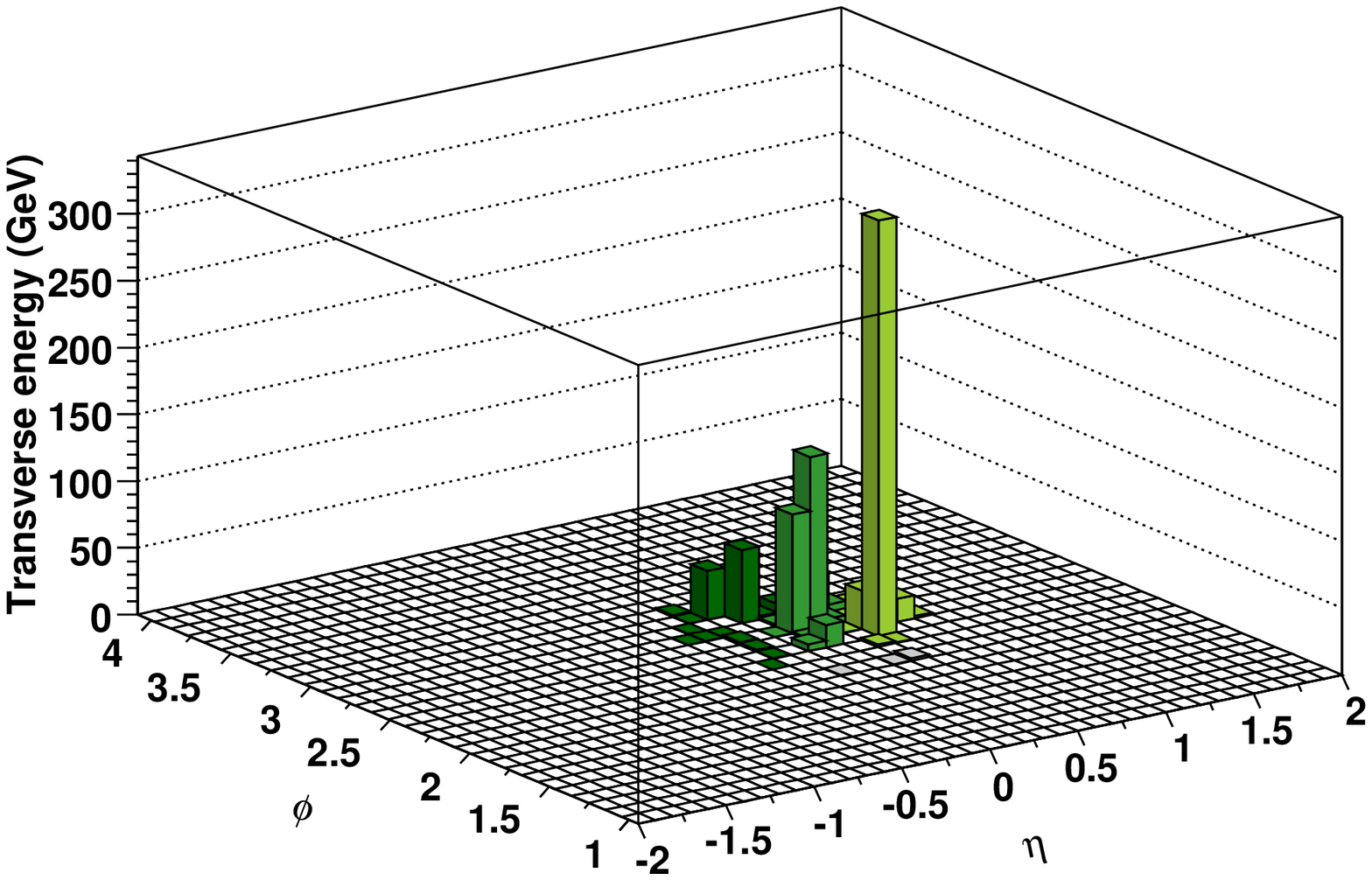}}\\
\emph{{Fig.1.}} {\emph{Top jet with a $p_T$ of 800 GeV at the LHC.}}
\ec
	
	Monte Carlo simulations are used for b-tagging connected with experimental studying of modern particle physics. The b-tagging tool has the following components:\\

    a) \textbf{\textit{Algorithms for jet identification}} 
    
	Jets are the sprays of hadrons from the fragmentation of quarks or gluons and can be identified by measuring energy and direction. So one should have rules for the  projection of a set of particles onto a set of jets. Such a set of rules is referred as a jet reconstruction algorithm, which together with its parameters is known as a jet definition. 
Jet reconstruction is made from the QCD-multijet event sample with the corresponding algorithm with a cone size $\Delta R$ in the $(\eta, \phi)$ plane, from several types of inputs, \cite{6.}.

	Jet clustering algorithms are among the main tools for analysing data from hadronic collisions. Their widespread use at the LHC have stimulated considerable debate concerning the merits of different kinds of jet algorithm. Among them are sequential recombination algorithms, $k_t$ and so called anti-$k_t$. The extension relative to the $k_t$ and anti-$k_t$ algorithms lies in a definition of the distance measures:\\
For each pair of particles $i, j$ work out the $k_t$ distance
\[d_{ij}=min(k^{2p}_{ti}, k^{2p}_{tj})\Delta R^2_{ij}/R^2 \ ,\]
\[d_{iB}=k^{2p}_{ti}.\]
with $\Delta R^2_{ij}=(y_i-y_j)^2+(\phi_i-\phi_j)^2$, where
$k_{ti},\ y_i$ and $\phi_i$ are the transverse momentum, rapidity 
and azimuth of particle $i$ and $R$ is a jet-radius parameter taken of order 1; for each parton $i$ also work out the beam distance, $d_{iB}=k^2_{ti}$.
For $p = 1$ one recovers the inclusive $k_t$ algorithm. 
The case of $p = 0$ is special and it corresponds to the
inclusive Cambridge/Aachen algorithm.
Negative values of $p$ might at first sight seem pathological. The behaviour with respect to soft radiation will be similar for all $p < 0$, so it is common to use $p = −1$, and refer to it as the "anti-$k_t$” jet-clustering algorithm.

	b) \textbf{\textit{Impact parameter identification}}

	For increasing the fraction of well-reconstructed tracks and for reduction of the contamination by long-lived particles is used selection on the track impact parameters. The impact parameters dxy and dz are defined as the transverse and longitudinal distance to the primary vertex at the point of closest approach in the transverse plane. Representation of impact parameter is in fig. 2
\bec 
{\includegraphics[width=0.42\textwidth]{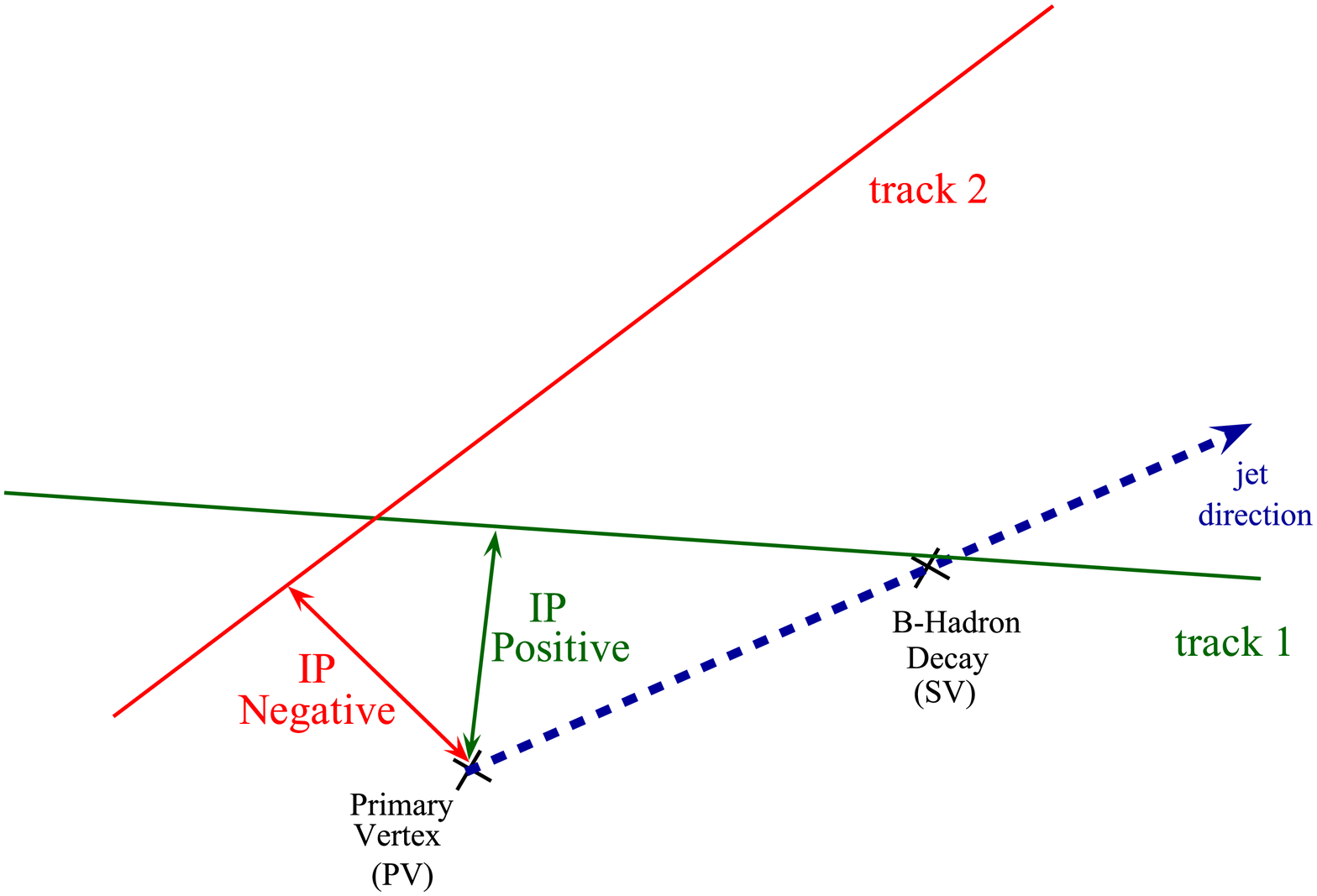}}\\
\emph{{Fig.2.}} {\emph{Definition of positive and negative impact parameter
s.}}
\ec

	Tracks are associated to jets in a cone  $\Delta R=\sqrt{(\Delta\phi)^2+(\Delta\eta)^2}$. The pseudorapidity $\eta = ln[tan(\theta /2)]$ is defined through the polar angle,$\theta$, measured from the positive z axis and through the azimuthal angle, $\phi$, measured in the x-y plane. The point of closest approach must be within 5 cm of the primary vertex. Such discrimination of impact parameters is the basis for all our calculations. In \cite{7.} are presented the most critical track parameters for b-tagging - the transverse impact parameters as the distance of closest approach of the track to the primary vertex point in the $r\phi$ projection, fig.3. 	
\bec 
{\includegraphics[width=0.59\textwidth]{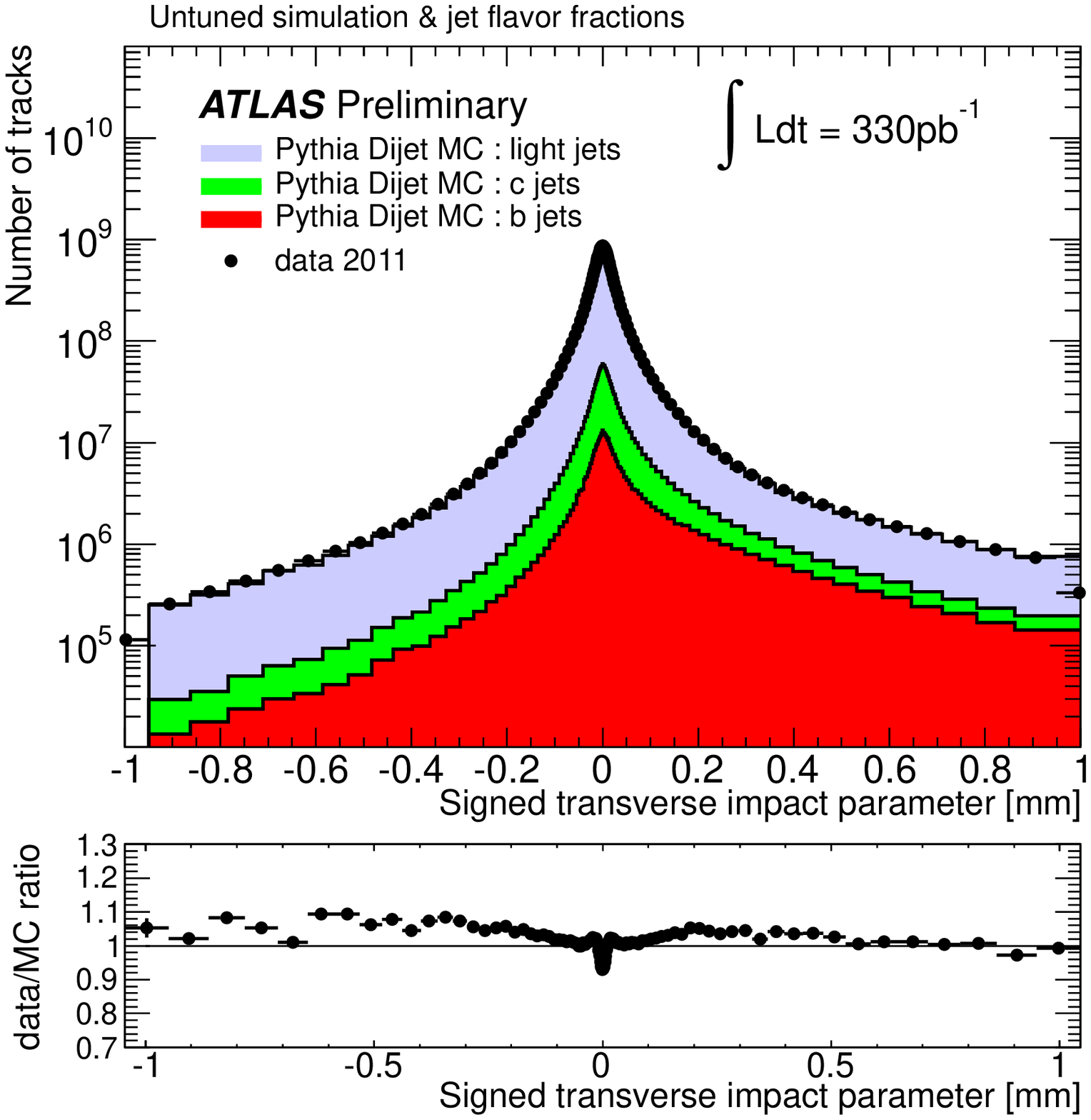}}\\
\emph{{Fig.3.}} {\emph{Distribution of the signed transverse impact parameter $d_0$ with respect to primary vertex for tracks of b-tagging quality associated to jets, for experimental data (solid black points) and for simulated data (filled histograms.}}
\ec

	    c) \textbf{\textit{Secondary vertex identification}}
	    
	For data selection between b and non-b jets can be used the presence of a secondary vertex identification. Secondary-vertex candidates must meet the following requirements to enhance the b purity:

	$\bullet$ secondary vertex candidates with a radial distance of more than 5 cm with respect to the primary vertex, with masses compatible with the mass of $K^0$ or exceeding 6.5 GeV/c$^2$ are rejected;
	
	$\bullet$ b-quarks/B-hadrons have sizeable lifetime, 1.6 ps, which corresponds to c$\tau\sim$ 500 micro meters;
    
	$\bullet$ the flight direction of each candidate has to be within a cone of $\Delta R <$  0.5 around the jet direction.
	
	The rejection of non b-jets is dependent on the fact that light quarks are connected with stable particles and appear from the primary vertex. The extension is connected with the impact parameter resolution connected with long lifetimes of b-hadrons. Therefore, b-jets have large excess of primary vertex corresponding to genuine lifetime content. In \cite{8.} is presented the secondary vertex reconstruction efficiency as function of jet $p_T$ and $\eta$ for b-, c- and light-flavour jets, fig. 4 . 
\bec 
{\includegraphics[width=0.39\textwidth]{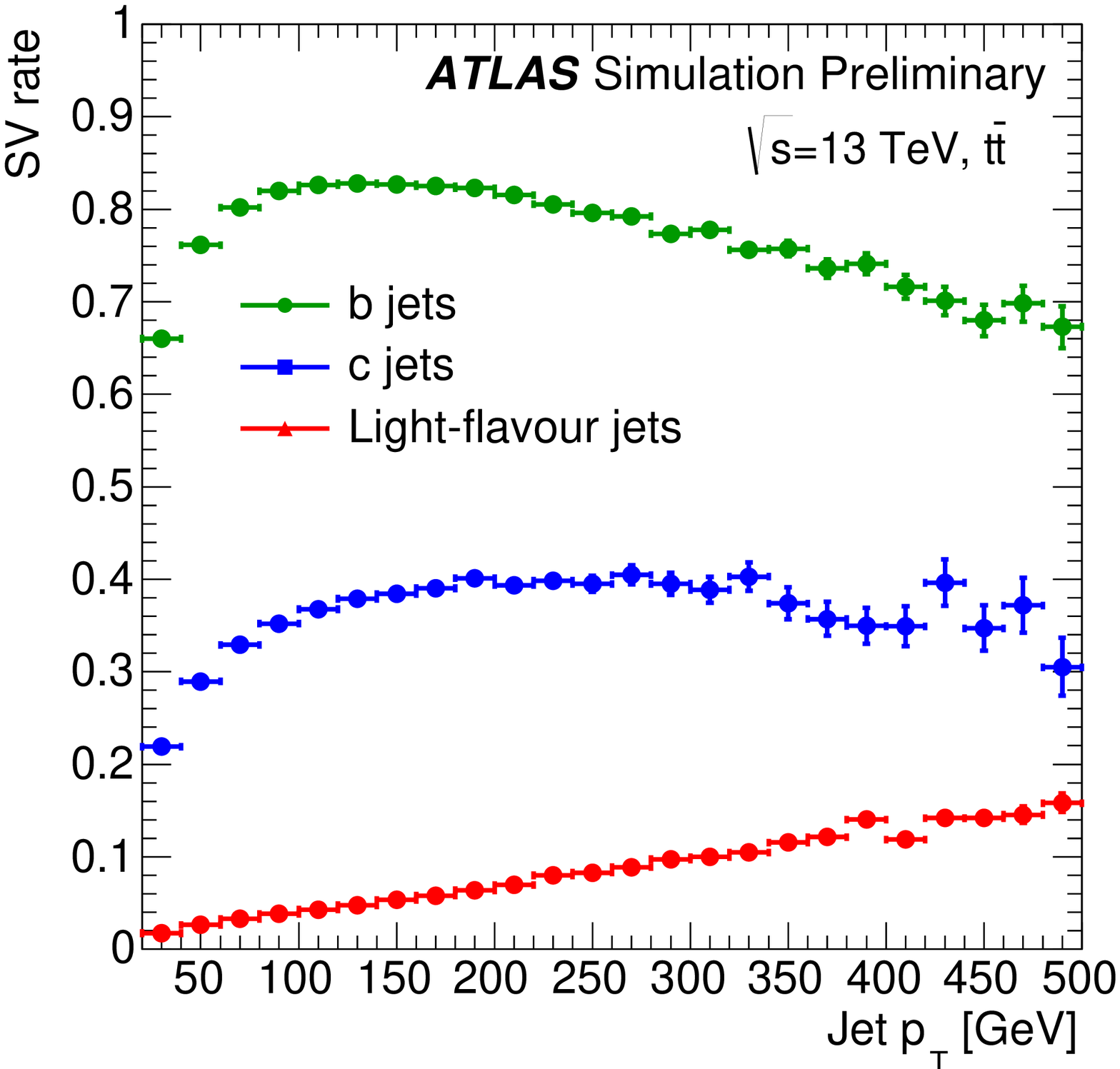}}
{\includegraphics[width=0.39\textwidth]{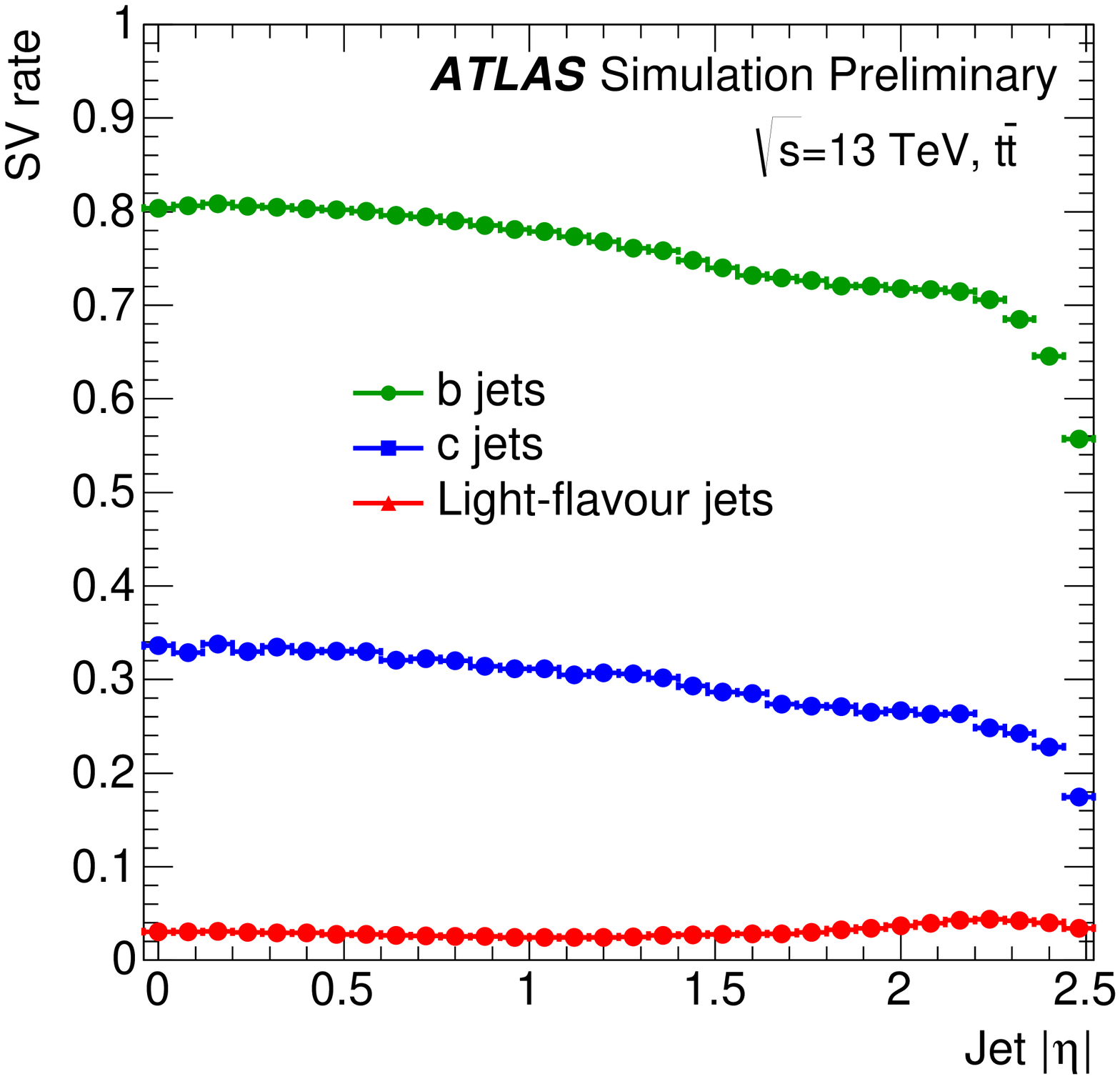}}\\
\emph{{Fig.4.}} {\emph{Secondary vertex reconstruction rate as function of jet $p_T$ (left) and $\eta$ (right) for b (green), c (blue) and light-flavour (red) jets in t-anti t-events.}}
\ec

	d) \textbf{\textit{Kinematic cuts}}
	
	Kinematic cuts are connected with physical processes for searches of new physics and with geometry of experimental setup of LHC, FCC or ILC type. 
This channel is of interest because of top Yukawa coupling connected with problems of vacuum stability in the Standard Model. As this reaction is limited by experimental systematics and theory uncertainties it can be studied at the FCC collider \cite{9.} with the increased production cross section to 
\[\sigma(pp\rightarrow t\bar{t}h, 100 TeV)=\ 33.9\  pb\ .\]
From \cite{10.} we can take provided predictions for the inclusive jet cross sections. The jet $p_T$ threshold was taken from 50 GeV/c to 100 GeV/c and $\eta <$ 2.5. Jets were defined using anti-k$_T$ algorithm with $\Delta R$=0.4.

	e) \textbf{\textit{Computer modelling}}

	Monte Carlo (MC) simulated samples of multijet events were generated
with Pythia 6.42 and Pythia 8.2, \cite{11.}. Then we have used Fastjet 3.3.0, \cite{6.}, program for jet identification. We also have used MadGraph program, 
\cite{12.}, for the extraction of the information about kinematic data (p$_T$, $\eta$ , y) and about energy distribution of the events.

\section{Results of calculations}
Using FastJet code (version 3.3.0) for jet identification with the help of computer program Pythya 8.2 and using kinematic data selected by experimental methods and computer simulation we carried out calculations on computer modelling of proton-proton collisions for the most interesting from a physical-theoretical point of view, processes of the formation of top quarks. For the future experiment at the LHC we have performed our calculations at the energy 14 TeV.
	
	We have calculated the number of events, N as the function of distance between track and jet axis, D, $dN/dD$, and found the best $D\sim$ 0.4 presented in fig 5. 
\bec 
{\includegraphics[width=0.5\textwidth]{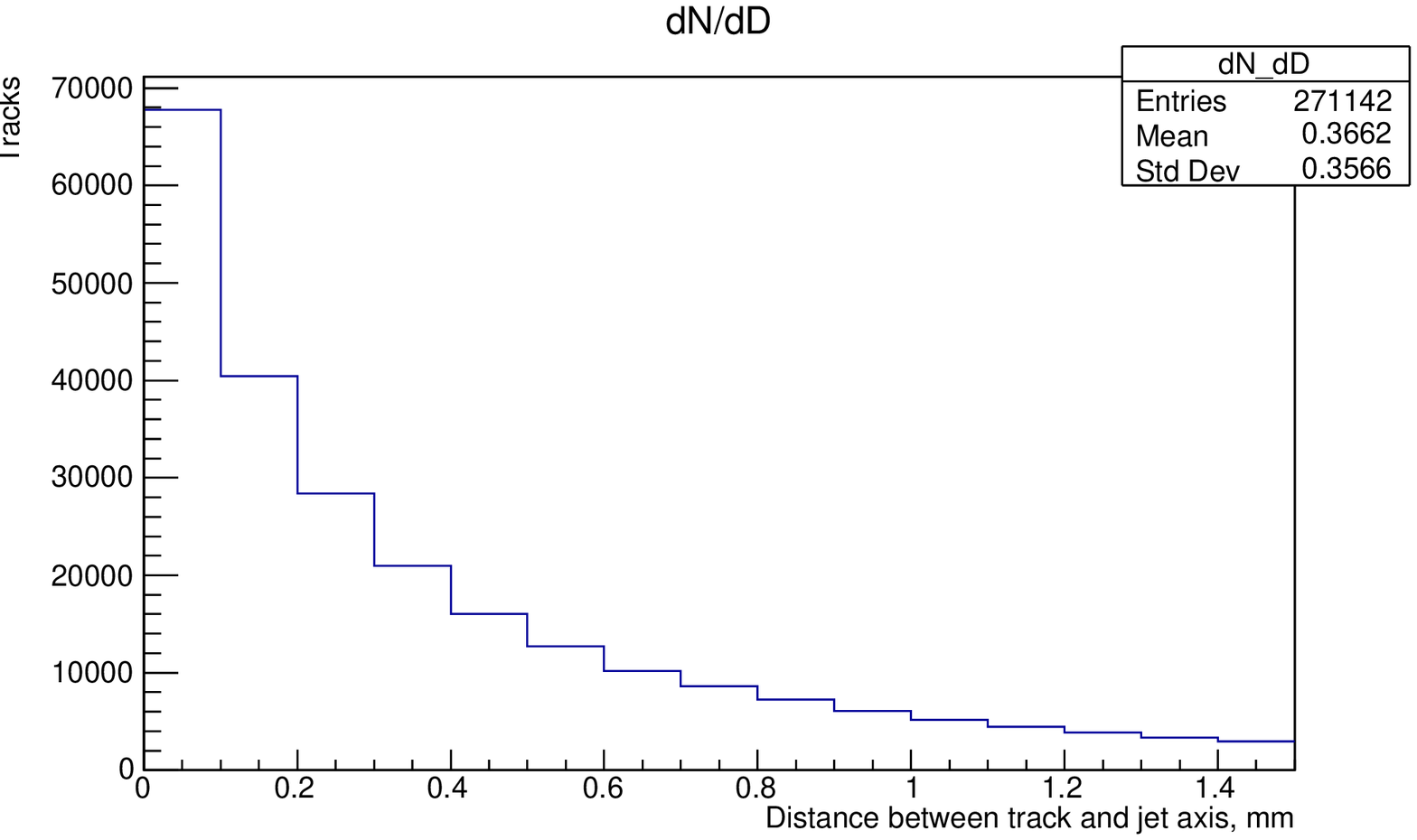}}\\
\emph{{Fig.5.}} {\emph{The  number of tracks, N as the function of distance between track and jet axis, D.}}
\ec
and the number of events, N as the function of transverse momentum, $p_T$, fig. 6 and found the best result for $p_T\sim$ 40-100 GeV
\bec 
{\includegraphics[width=0.5\textwidth]{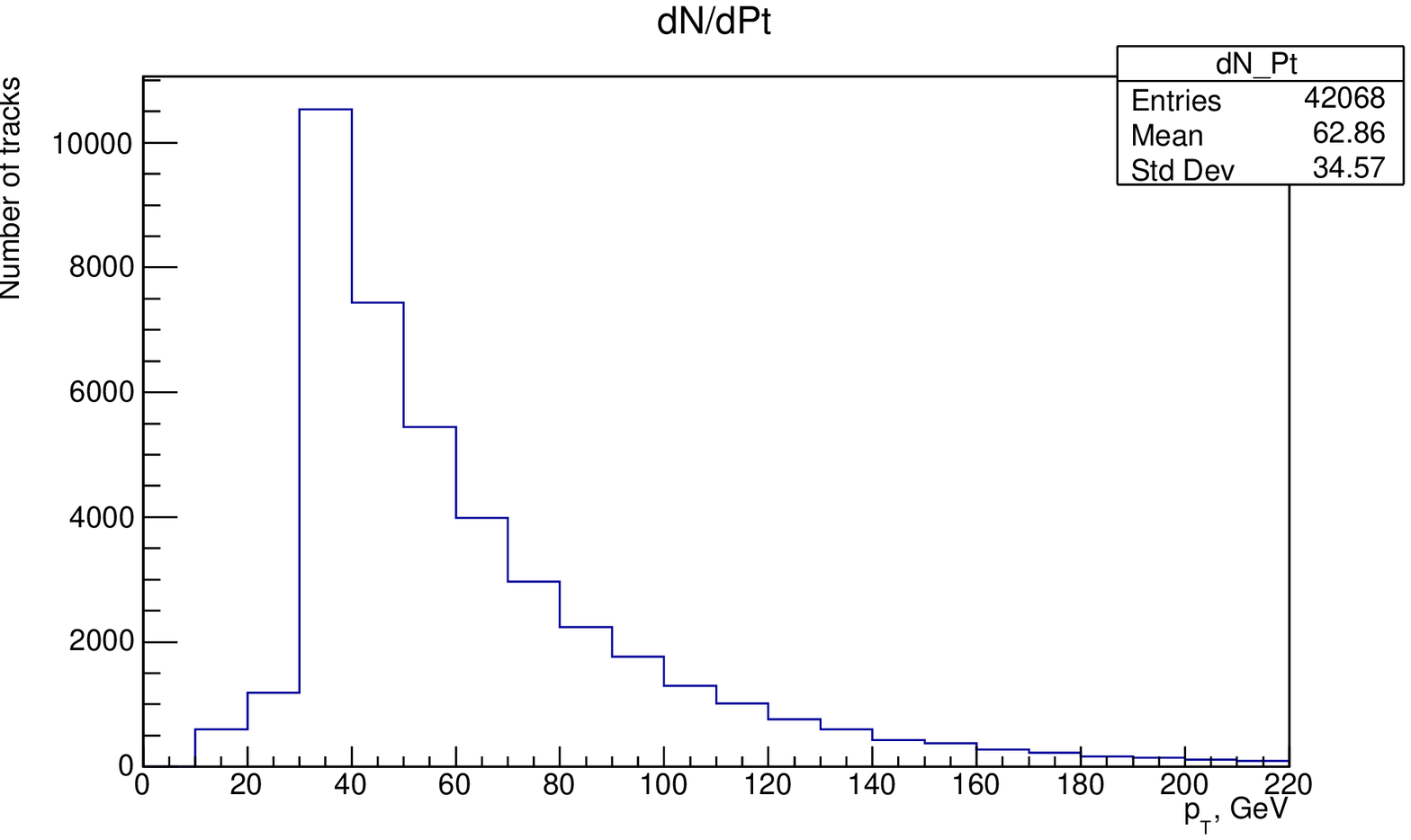}}\\
\emph{{Fig.6.}} {\emph{The number of tracks, N as the function of transverse momentum, $p_T$.}}
\ec
In fig. 7 is presented the number of events, N as the function of rapidity, $y$.
\bec 
{\includegraphics[width=0.5\textwidth]{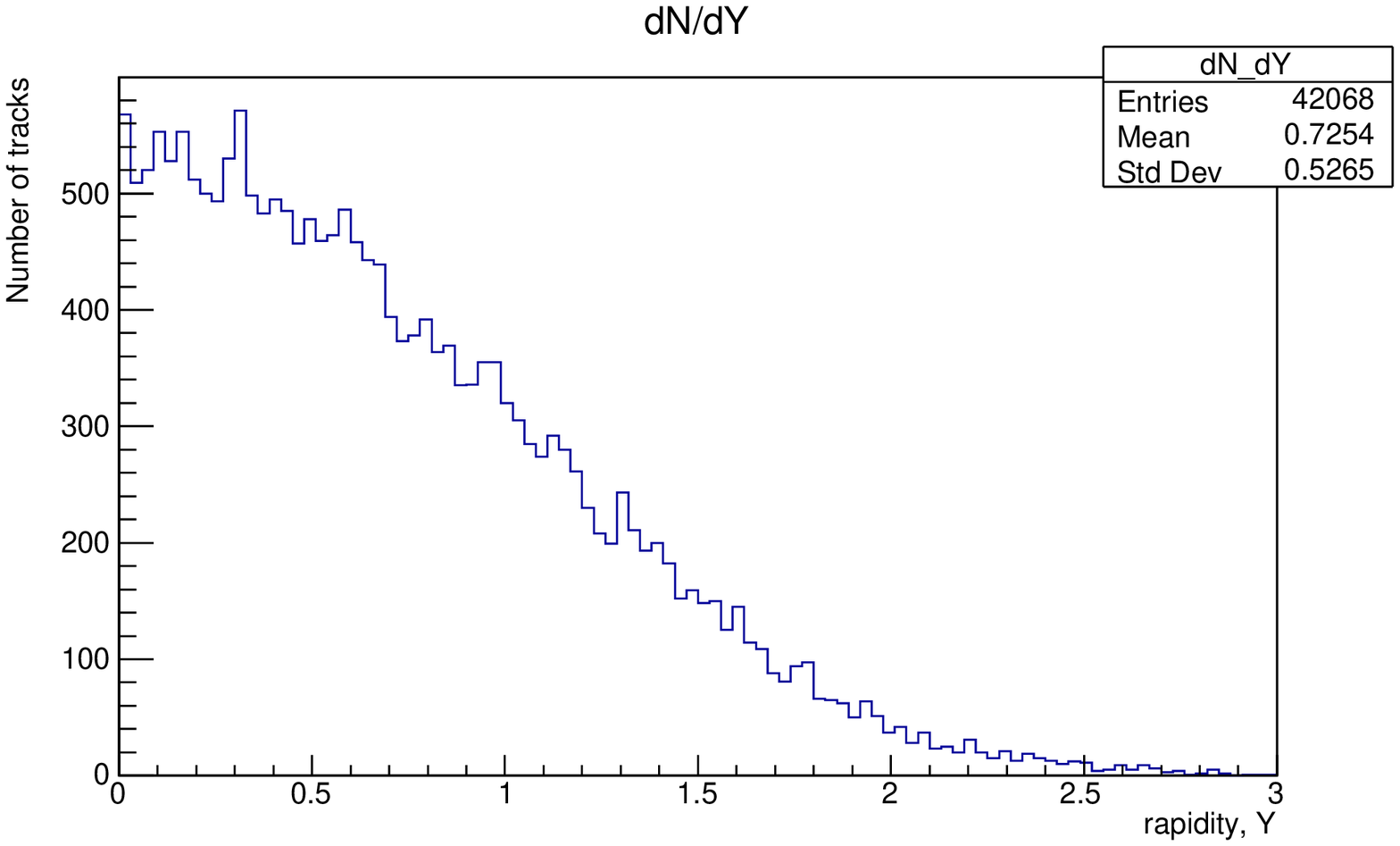}}\\
\emph{{Fig.7.}} {\emph{The number of tracks, N as the function of rapidity, $y$.}}
\ec

In fig.8 is presented process $gg\rightarrow Ht\bar{t}$ for $p_T >$ 50 GeV and $p_T>$ 100 GeV
\bec 
{\includegraphics[width=0.39\textwidth]{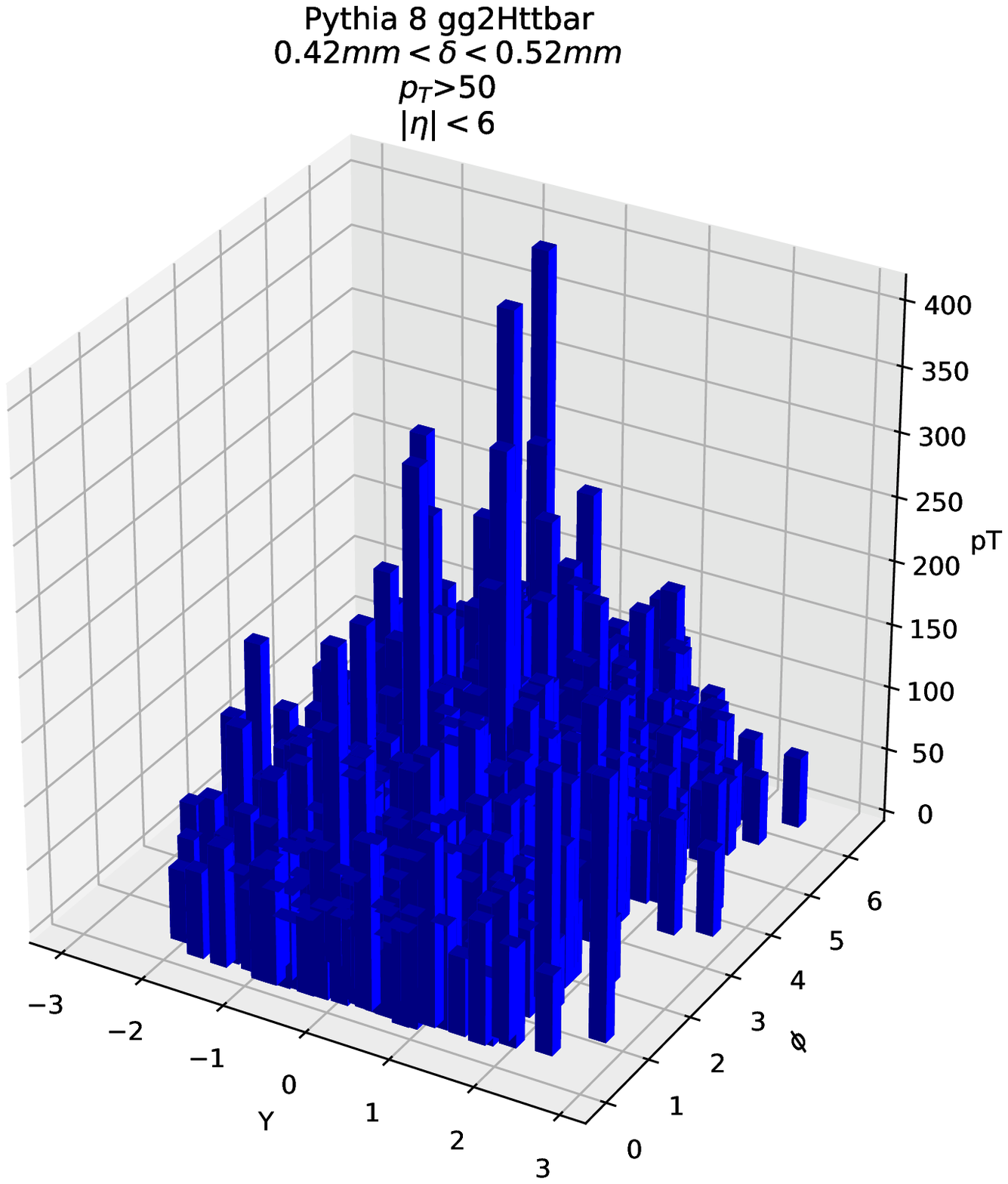}}
{\includegraphics[width=0.39\textwidth]{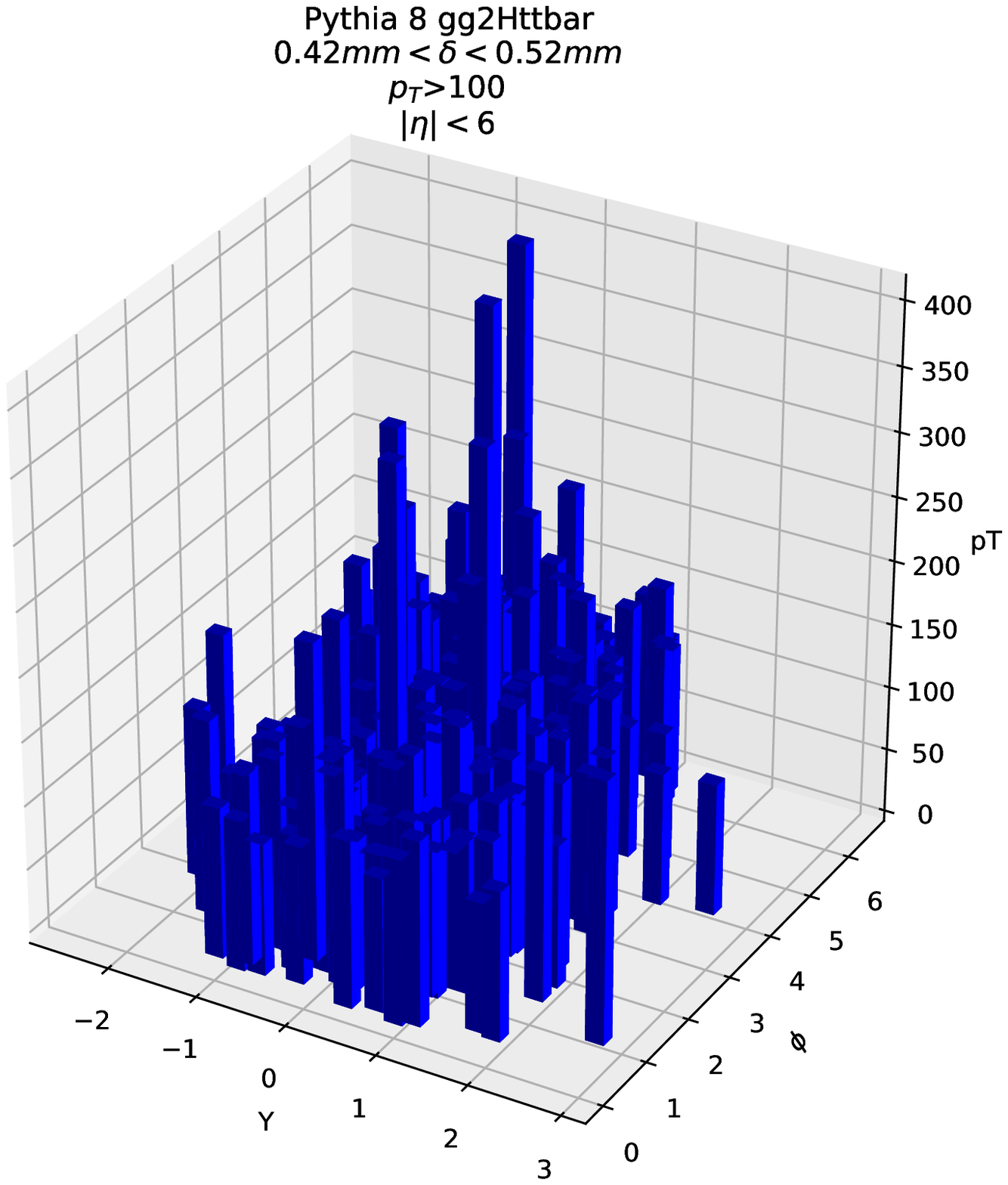}}\\
\emph{{Fig.8.}} {\emph{Process $gg\rightarrow Ht\bar{t}$ for $p_T >$ 50 GeV (left)  and $p_T>$ 100 GeV (right).}}
\ec
From the comparision of these two fig. 8 we can see that the jets in the second case are less spread out over the plane of the angles and are more high-energy.
In fig. 9 are compared two processes $gg\rightarrow t\bar{t}$ and $qq\rightarrow t\bar{t}$ with the same cinematical cuts
\bec 
{\includegraphics[width=0.39\textwidth]{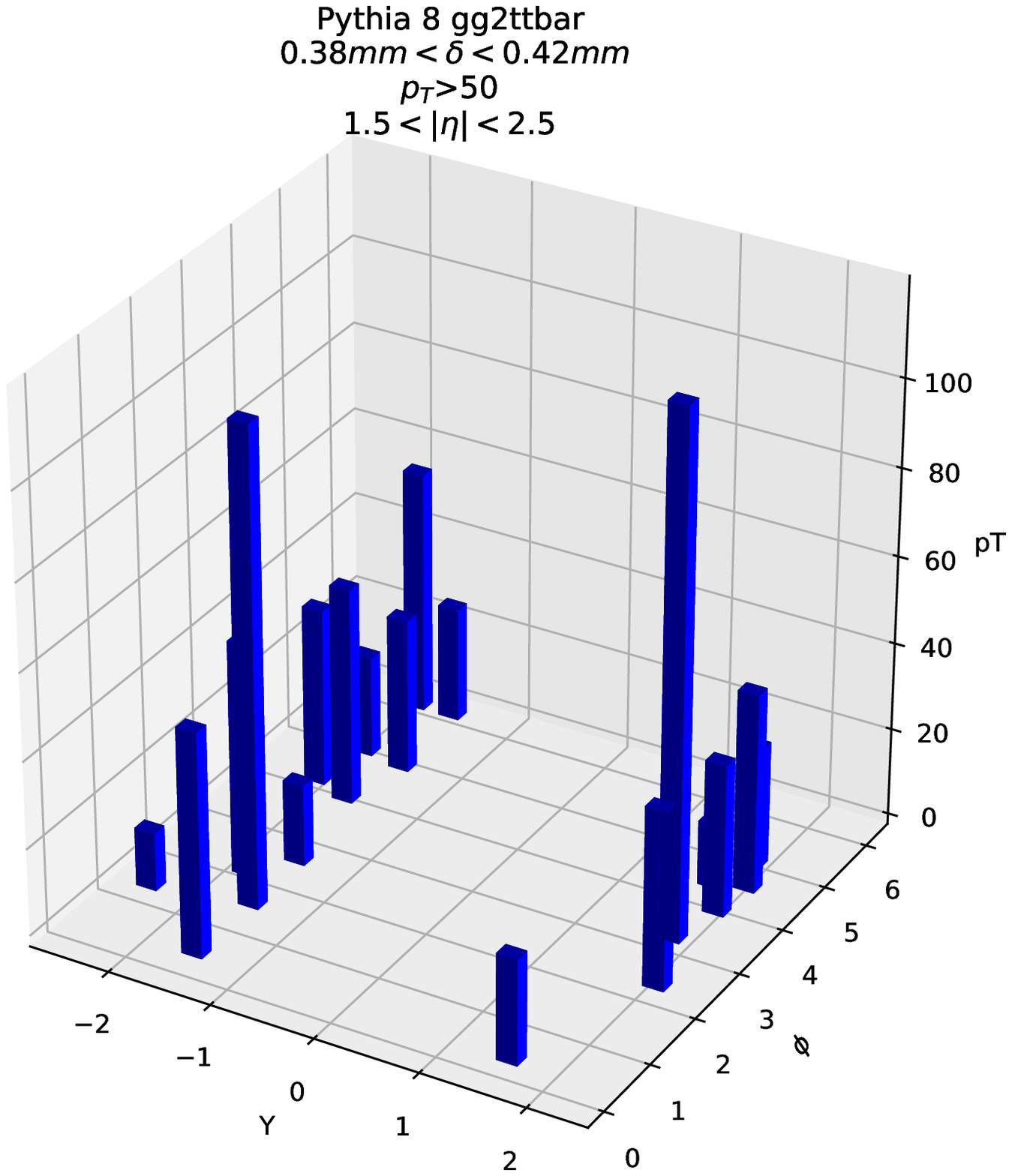}}
{\includegraphics[width=0.39\textwidth]{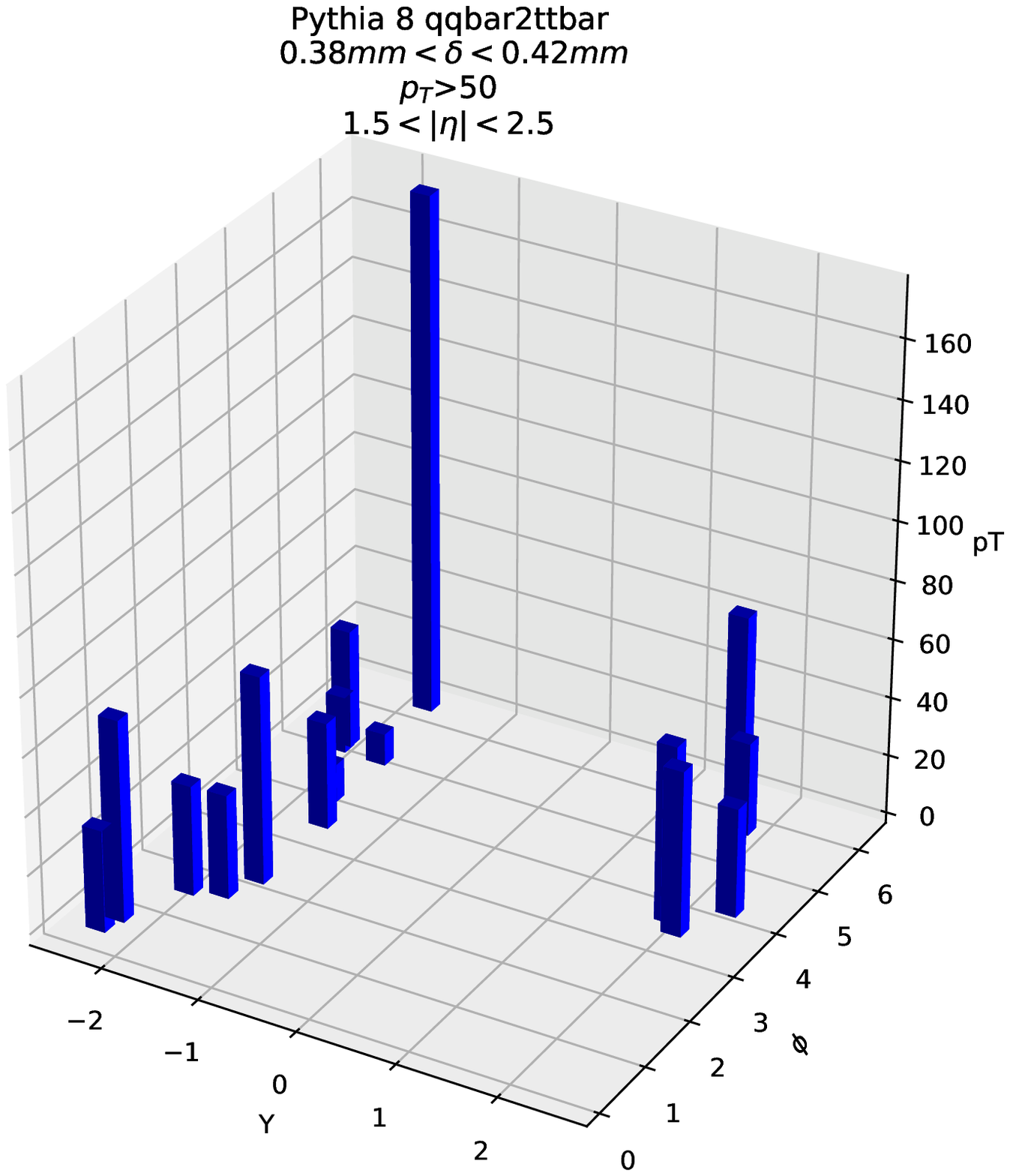}}\\
\emph{{Fig.9.}} {\emph{Process $gg\rightarrow t\bar{t}$ for $p_T >$ 50 GeV (left)  and $qq\rightarrow t\bar{t}$ for $p_T >$ 50 GeV (right).}}
\ec
The number of jets for the $gg\rightarrow t\bar{t}$ process is much more than for the  $qq\rightarrow t\bar{t}$ process with the same kinematic data.

	In fig. 10 are presented two processes $gg\rightarrow t\bar{t}$ and $qq\rightarrow t\bar{t}$ with another cinematical cuts on the impact parameter, and $p_T$
\bec 
{\includegraphics[width=0.39\textwidth]{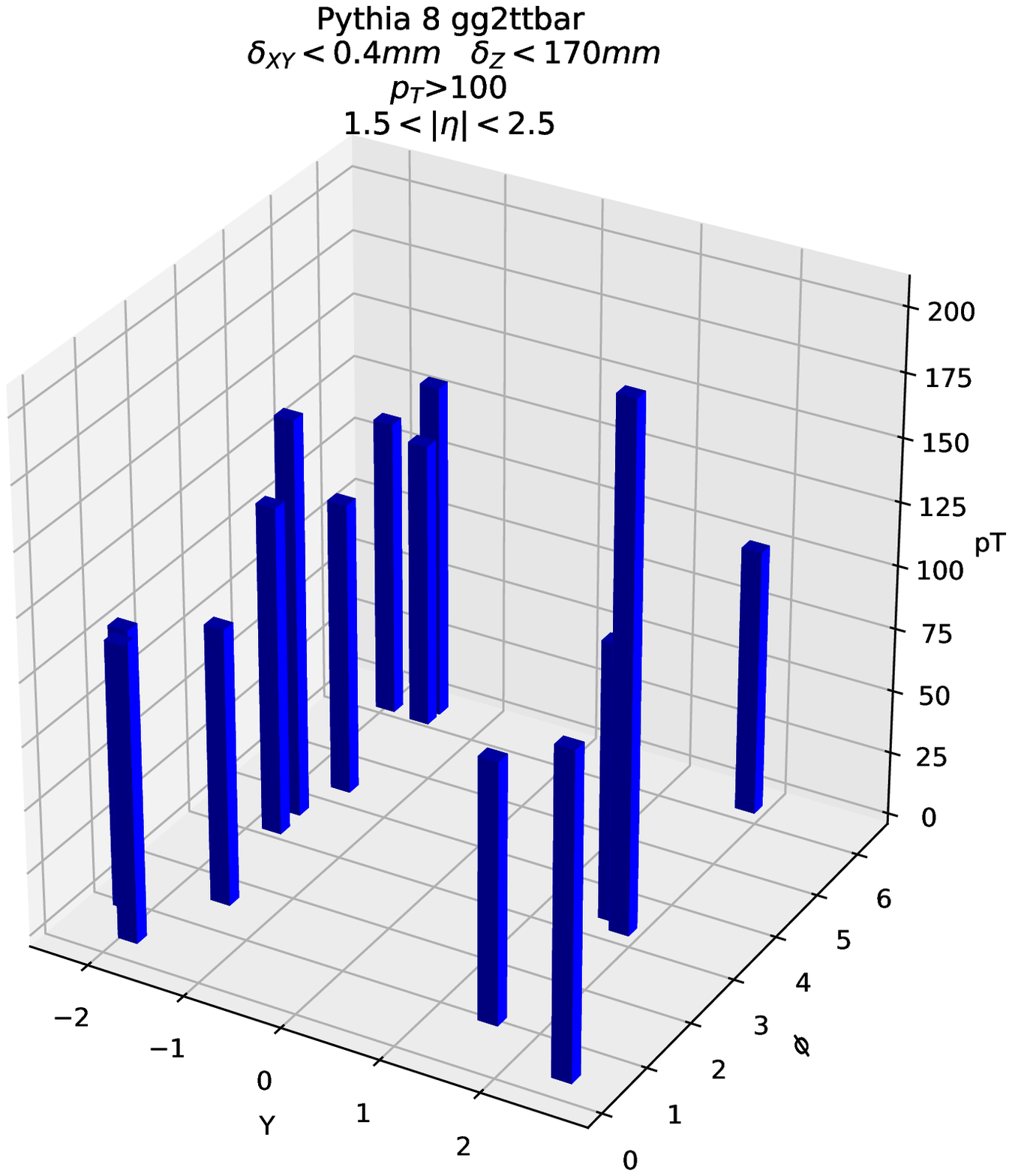}}
{\includegraphics[width=0.39\textwidth]{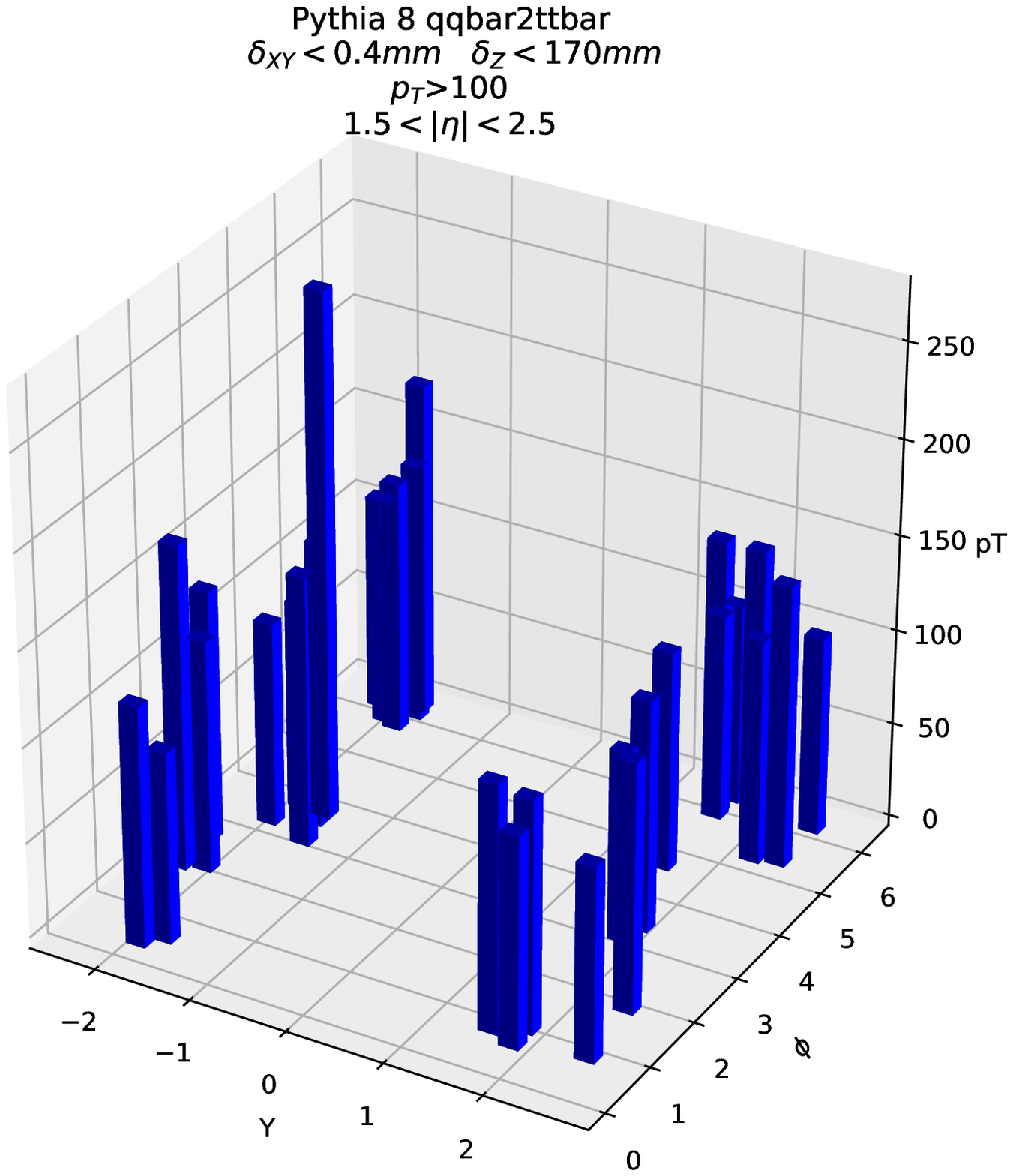}}\\
\emph{{Fig.10.}} {\emph{Process $gg\rightarrow t\bar{t}$ for $p_T >$ 100 GeV (left)  and $qq\rightarrow t\bar{t}$ for $p_T >$ 100 GeV (right).}}
\ec
From comparison of these figures we can say about the predominance of the number of jets for $qq\rightarrow t\bar{t}$ compared with $gg\rightarrow t\bar{t}$ for $p_T >$ 100 GeV at the impact parameter data $\delta_{xy}<$ 0.4 mm and $\delta_z <$ 170 mm.

\section{Conclusion}
We have performed the computer modeling of the proton-proton collisions with top quarks in the final state at the energy of 14 TeV at the LHC for the following processes:
	
	$gg\rightarrow Ht\bar{t}$;
	
	$gg\rightarrow t\bar{t}$;
	
	$qq\rightarrow t\bar{t}$.
	
Our calculations were performed with different kinematic cuts on:

	$p_T$;
	
	$\delta_{xy}$;
	
	$\eta$.

From our calculations we came to the following conclusions for the larger number of jets:

	$\bullet$ the predominance of the gg processes compared to qq processes for $p_T >$ 50 GeV and $0.38 mm < \delta_{xy}<0.42 mm$;
	
	$\bullet$ the predominance of the qq processes compared to gg processes for $p_T >$ 100 GeV and $\delta_{xy}<0.4$ mm , $\delta_{z}<170$ mm.

We also made a conclusion about the decrease of jet dispersion in the angle plane ($y$ and $\phi$) and transverse momentum with increasing boundary on the transverse momentum from $p_T>$ 50 GeV to $p_T>$ 100 Gev in the reaction $gg\rightarrow Ht\bar{t}$. 

    We clarified the predominance of such kinematic cuts as:
     
    1. impact parameter;
    
    2. transverse momentum;
    
    3. weak dependence on delta R;
    
    4. the relationship of the magnitude of the transverse momentum and angular variables.

\section{Conflicts of Interest}
The authors declare that there is no conflict of interest
regarding the publication of this paper.

\label{page-last} 
\label{last-page}
\end{document}